\documentstyle[12pt]{article}
\begin{document}
\begin{titlepage}
\rightline{\vbox{\halign{&#\hfil\cr
&SWAT/35\cr
&\today\cr}}}
\vspace{0.5in}
\begin{center}
{\Large\bf
Simulation of Field Theories in Wavelet Representation}\\
\medskip
\vskip0.5in

\normalsize {I.G. Halliday and P. Suranyi\footnote{Permanent address:
Department of Physics, University of Cincinnati, Cincinnati, Ohio, 45221
U.S.A.}
\smallskip
\medskip

{ \sl Department of Physics, University of Wales}\\
Swansea, Wales, United Kingdom SA2 8PP\\ \smallskip}
\end{center}
\vskip1.0in

\begin{abstract}
The field is expanded in a wavelet series and the wavelet coefficients
are varied in
a simulation of the 2D $\phi^4$ field theory. The drastically reduced
autocorrelations
result in a substantial decrease of computing requirements, compared to
those in
local Metropolis simulations. The improvement
is shown to be the result of an additional freedom in the choice of the
allowed range
of change at the Metropolis update of wavelet components, namely the range
can be
optimized independently for all wavelet sizes.
\end{abstract}
\end{titlepage}
\section{Introduction}

In the present work we introduce
 a new approach
to simulations of field theories, based on wavelet expansions.
{}~\cite{debuch}~\cite{book}
{}~\cite{c-book}
Wavelets form  complete sets of localized, orthonormal states. Elements
of a set of
wavelets differ both in their
locations and in their scales. In fact, complete sets contain wavelets
at all scales, starting from an elementary scale up to the scale of the
system. Wavelets
can be labeled by two integers, $n$ and $k$, characterizing their scale
and their location.
 They can be formed from a function $\psi(x)$ as
 \begin{equation}
 \psi_{k,n}(x)=2^{-n/2}\psi(2^nx+k).
 \label{wave_gen}
 \end{equation}

Wavelets have been applied succcessfully to linear problems, like
signal analysis.
In the continuum, wavelets form a  complete orthonormal system,
allowing the
expansion of $L^2$ integrable functions, in a manner analogous
 to Fourier expansion.
The orthogonal functions of the Fourier expansion are not
localized, consequently they are not normalizable. Wavelets
are localized, albeit on varying scales, from the
elementary scale to the scale of the system. They are themselves
$L^2$ integrable function. As a rule,
they lead to series expansions with convergence considerably faster
then that of
a Fourier expansion.

  The property of wavelets of analysing data at all scales makes them,
  in principle,
 attractive for
the investigation of lattice problems. One can define wavelet
representations,
in which fields are expressed by their wavelet expansion coefficients,
rather then by their values at given spacetime points. On finite lattices,
the range of
integers $k$ and $n$ in (\ref{wave_gen}) is finite. It is not obvious
that wavelet
expansions, designed for investigating linear problems are useful for
investigating
a highly nonlinear problem such as a lattice field theory. The presence
of large scale
wavelets introduces nonlocality in the expression for interaction terms.
The decrese
in autocorrelation time is partially counterbalanced by the increased
computational
requirements. It is then important to give estimates for the growth
of nonlinearity
in terms of the lattice size. The question is the balance:
 whether one gains more by decreasing autocorrelation times
 than one loses by having to deal with a number of nonlocal terms
in the action.  This question will be investigated extensively in
the remainder of
this paper.
 Wavelets have already been used in lattice problems:  as a
 variational basis in the
$XY$ model~\cite{best}, and as a tool to facilitate gauge
fixing~\cite{what}.

Simulations in wavelet representation are somewhat similar to simulations
using collective updating methods, such as Swendsen and Wang's
algorithm~\cite{swendsen} or
the multigrid method~\cite{brandt}.
While the multigrid method is a so called smoothing
filter, averaging a large number of points, the wavelet representation
is, in a way, the opposite. Wavelets are designed to have their first few
moments vanish. Collective updating methods have been designed to
decrease the
dynamical exponent $z$ appearing in the expression of the autocorrelation
time ($\tau$) of physical quantities
\begin{equation}
\tau\simeq c ~\xi^z,
\label{auto}
\end{equation}
where $\xi$ is the spacetime correlation length, the inverse of the
mass gap.
In fact, some of the algorithms are able to reduce $z$ to almost zero.
 Near second order phase transition points the correlation length is
 proportional
to the linear size of the lattice, $L$. Thus, collective updating
algorithms are
particulary useful near critical
points and for large lattices. In fact, for lattices of small or
moderate size
or at some distance away from the critical point,  most collective
updating algorithms
are not much superior to local updating methods.

 In the next section we will introduce the wavelet formalism
 and transform the $\phi^4$ field theory into wavelet representation.
 Then we will count the number of nonlocal terms,
  to get a handle on the increase of
 computational needs as a
 function of lattice size.  In Section 3. we put the wavelet method
 to a direct test, we simulate a 2 dimensional $\phi^4$ theory, both
 using the
 standard Metropolis algorithm and the wavelet method. The last section
 contains our
 conclusions.

\section{The Wavelet Method in Field Theories}
Wavelets form a orthonormal system of variables interpolating between
coordinate and
momentum representations. In fact, wavelet representations (there are many
different choices) unify the advantages of both. They are much more local
and stable
against small perturbations than the momentum representation, and
at the same
time they can describe long range correlations with a small number of terms,
unlike the coordinate representation.

The most widely used wavelet filters were discovered by Daubechies
{}~\cite{debuch}.
Daubechies wavelets on discrete sets are defined as orthogonal matrix
transforms. 2$p$ of $2^n$ spatial
components, $x_1,...,x_{2^n}$ are combined together in two different ways:
with coefficients
 $c_1x_1+c_2x_2+...+c_{2p}x_{2p}$ (smoothed combinations)
 and with different ordering and signs of the same
 coefficients $-c_{2p}x_1+c_{2p-1}x_2-...+c_1x_{2p}$ (wavelets). The
 series is required
to have $p$ vanishing moments, i.e. $\sum_k (-k)^lc_{2p-k}=0,~
l=0,1,...,p-1$. Similarly,
one forms such linear combination from subsequent components $x_i$,
shifted by two
lattice
units an arbitrary number of times. Then requiring the orthogonality
of the
transformation makes the coefficients
essentially unique. The wavelet transform consists of $n-p$ subsequent
orthogonal
transformations
of the above described type, applied to the smoothed combinations obtained
in the previous orthogonal transformation. The coefficients
$c_1,...,c_{2p}$
 are called the Daubechies wavelet
filter coefficients.

For $p=1$  $c_0=c_1=1/\sqrt{2}$, and the wavelets are called
Haar-wavelets~\cite{book}.
The $r$th Haar wavelet of length $2^k$ on a one dimensional
lattice is
formed as
\begin{equation}
\chi_{k,r}=2^{-k/2}\sum_{i=1}^{2^{k-1}}[\phi(i+r 2^k)-\phi(i+
2^{k-1}+r 2^k)],
\label{wavelet}
\end{equation}
where $\phi(j)$ denotes the field at the $n$th site. The range
of $k$ and $r$ is
$k=1,2,...,n$, and $r=0,...,2^{n-k}-1$, respectively, where $L=2^n$,
is the size
of the lattice.
The orthogonal system is formed from the wavelets $\chi(k,r)$,
and one "smoothed" combination
\begin{equation}
\chi=\frac{1}{\sqrt{L}}\sum_{i=1}^L \phi(i),
\label{magn}
\end{equation}
which is the normalized total magnetization.

 The structure of the wavelet coefficients for general $p$ is
 similar to
 that of (\ref{wavelet}). Thus, the first $2^{n-1}$ wavelets are
 combinations
 of $2p$ neighboring data points shifted by multiples of 2.
 Then the next $2^{n-2}$ wavelets (each shifted from the next by 4
 data points)
  are of length $l_2=1+(2^2-1)(2p-1)$, etc. The $2^{n-k-1}$ $k$th
  type of wavelets are
  of length $l_k=1+(2^k-1)(2p-1)$. They are shifted from each other
  by $2^k$ points.
  The last $p$ combinations are not wavelets, but maximally smoothed
  combinations
  of the original data points.

 In an application to lattice field theory one has to use wavelets in
 more
 then one dimension.  In fact, wavelets can easily be defined in
 any number of dimensions.
In each dimension the wavelets are labeled by their scale,
$k_i$, and by their location, {\bf r}. Thus, a wavelet coefficient
$\chi(k_1,r_1;k_2,r_2;...;
k_d,r_d)$ corresponds to a  wavelet box of length $2^{k_i}$ in the
$i$th lattice dimension.
$d$ is the number
of dimensions.
 There are $2^{nd-k_1-k_2-...-k_d}$
wavelets labeled by a given scale vector ${\bf k}=\{k_1,k_2,...,k_d\}
$. The lattice field has a linear expansion in terms of the coefficients
$\chi(k_1,r_1;k_2,r_2;...;
k_d,r_d)$. We will call these coefficients the wavelet representation.
It is on an equal
footing with the momentum and coordinate representations.

Wavelets are well suited to analyzing linear problems. Consequently
a Gaussian
model is just as simple in wavelet representation as in coordinate
representation.
Orthonormality implies that the form of the mass term
 is unchanged in wavelet representation
\begin{equation}
S=\frac{m^2}{2}\sum_{j,r}2|\chi(j,r)|^2
\label{newaction}
\end{equation}
where for the sake of simplicity we have dropped the limits of
the summations.
The kinetic term is slightly more complicated, and its explicit form is not
very enlightening.
 It has a moderate number of nonlocal terms.

The propagator in wavelet space is quite
analogous to the one in coordinate space. Thus, it has the same
asympotic behavior
and it also becomes singular at the critical point, $m^2=0$. Critical
phenomena can be
investigated in wavelet representation just as well as in coordinate
representation. Alternatively, one can easily transform correlation
functions
back to coordinate representation, using the wavelet transformation.

The question arises, however, how well the wavelet representation
is suited to
investigate nonlinear problems, like any nontrivial field theory.
In the current
paper we will
investigate a typical example
of nonlinear field theories, the $\phi^4$ theory. We need to
express the
interaction term in terms of wavelets. Using the orthogonality
of the wavelet
transformation an explicit
expression for  the interaction  term of the Lagrangian $g\sum
\phi^4$ is very
complicated.  It has many
non-local terms, some of them correspond to `long-range' interactions. The
success of the wavelet representation depends on the average number of
 these long range interaction terms per wavelet.  If this average number
increases slowly with size we have a chance that simulations in
wavelet representation will be more efficient then those in coordinate
representation.

First, we will calculate the average number of interaction terms per
wavelet
coefficient on a $d$-dimensional lattice. It is sufficient to do the
calculation in
one dimension. If the number of interaction terms per wavelet is $N$ in
one dimension then it is $N^d$ in $d$ dimensions. The interaction term
is a quartic combination of the wavelet coefficients. For the sake of
simplicity
we will do the counting for the Haar-wavelet representation. Then it
is easy to
see that in every term the coefficient with the lowest value of $j$
must be squared.
Let that coefficient be $\chi(j,m)$. Then the combinations of other
 wavelets multiplying $\chi(j,m)^2$ can be found as follows: for
 each $j$
there is only one wavelet  overlaping $\chi(j,m)$. Then the bilinear
multiplier
has $(n-j)(n-j+1)/2$ different types of terms. The number of wavelets
with scale $j$
is $L/2^j$. Then the total number of types of non-local interaction
terms is
\begin{equation}
LN=\sum_{j=1}^{n-1}\frac{L}{2^j}\ \frac{(n-j)(n-j+1)}{2}.
\label{number}
\end{equation}
In other words, the average number of non-local interaction terms
per wavelet is
\begin{equation}
N\simeq \frac{n^2}{2}=\frac{(\log L)^2}{2(\log2)^2}.
\label{Nest}
\end{equation}
This number should be compared with 1, the average number of terms per
site in the coordinate representation. In $d$ dimension the
appropriate factor is
\begin{equation}
N_d=\frac{(\log L)^{2d}}{2^d(\log2)^{2d}}.
 \label{Nd}
 \end{equation}

 In the next section, before discussing our simulations, we will perform
 a more precise analysis of computational needs.  It will turn out that
  the number of computations  per lattice site
 in a $d$ dimensional model increases only as  $(\log L)^d$ and not as
 $(\log L)^{2d}$. The application of multiple hits at each wavelet
 coefficient
 in the Monte Carlo simulation further decreases the ratio of computational
 needs.

 \section{Simulations}

 The difference in the number of cycles needed between local and
 wavelet simulations
 is due to the calculation of the change of the action.
 The largest computational needs arise from the calculation of the
 interaction term.
 The action has the form
 \begin{equation}
 S=\frac{1}{2}\sum_{\bf r}\left[m^2\phi_{\bf r}^2+
 \sum_j^d(\phi_{\bf r}-\phi_{\bf r-\hat
 {\bf e}_j})^2+\frac{g}{12} \phi_{\bf r}^4\right]
 \label{inter}
 \end{equation}
 This expression is very complicated in terms of the wavelet
 coefficients, even
 for the simplest, Haar wavelets. For that reason it is not
 conducive to the
 calculations to express (\ref{inter}) in terms of $\chi(k_1,r_1,
 ..., k_d,r_d)$.
 It is more efficient to store both the coordinate and the wavelet
 representation
 coefficients. This doubles storage needs, but accelerates and
 simplifies the code considerably.

 Suppose that during a sweep of the lattice we update the coefficient
 $\chi=\chi(k_1,r_1,..., k_d,r_d)$.
 Then
 the action has the following dependence on $\chi$:
 \begin{eqnarray}
 S&=&\frac{m^2\chi^2}{2}+\sum_{\bf r}\sum_j\left\{
 \chi[\phi_{\bf r}-\phi_{{\bf r}-\hat
 {\bf e}_j}]\left[
 \frac{\partial \phi_{\bf r}}{\partial \chi}
 -\frac{\partial \phi_{\bf r+\hat {\bf e}_j}}{\partial
 \chi}\right]+\frac{\chi^2}{2}\left[
 \frac{\partial \phi_{\bf r}}{\partial \chi}
 -\frac{\partial \phi_{\bf r+\hat {\bf e}_j}}{\partial
 \chi}\right]^2\right\}
 \nonumber\\&+&
 \frac{\chi^4}{4!}\sum_{\bf r}\left(\frac{\partial \phi_{\bf r}}
 {\partial \chi}\right)^4+\frac{\chi^3}{3!}\sum_{\bf r}\left(
 \frac{\partial \phi_{\bf r}}
 {\partial \chi}\right)^3\phi_{\bf r}+
\frac{\chi^2}{4}\sum_{\bf r}\left(\frac{\partial \phi_{\bf r}}
 {\partial \chi}\right)^2\phi^2_{\bf r}\nonumber\\&+&
 \frac{\chi}{6}\sum_{\bf r}\left(\frac{\partial \phi_{\bf r}}
 {\partial \chi}\right)\phi^3_{\bf r}+{\rm terms~independent~of~\chi},
 \label{action}
 \end{eqnarray}
  The
 coefficients
  $\partial \phi_{\bf r}/\partial\chi$ depend only on the
  relative position of
  the lattice points. They are given by products of coefficients
  $c_i^{(k)}$, for $k=
  k_1,k_2,...,k_d.$
These coefficients can be easily tabulated with minimal storage
requirements, even
for general $p$.  The
coefficient of $\chi^4$, and that of the quadratic part of the
kinetic term
 depend on ${\bf k}$ only. For Haar wavelets the coefficient of the
 fourth order
 term is
$2^{(k_1+k_2+...+k_d-dn)/2}$

The action defines the probability distribution of $\chi$. It must
be evaluated
repeatedly during simulations.The
calculation of the interaction part requires 8 floating point
operations (5 for Haar
wavelets) for every individual
coordinate point, ${\bf r}$. There are
$\prod l_{k_i}\simeq (2p-1)^d 2^{k_1+...+k_d}$ points
inside the wavelet. Now, for a complete sweep one has to update
all $2^{nd-k_1-
...-k_d}$ coefficients of the same kind, giving the number of
floating point operations
as $8(2p-1)^d2^{nd}$. Finally, one has to update the coefficients
for all values of
$k_1,...,k_d$, bringing in an additional coefficient of $n^d$,
where $n=\log L/\log 2$,
where $L$ is the length of an edge of the lattice. Then the
average number of
floating point operations per lattice site is
$N_d=8(2p-1)^dn^d$. The calculation of the kinetic term requires further
$4(2p-1)^dn^d$ operations. For Haar wavelets the calculation of the
kinetic term requires only  $O(n^{d-1})$
floating point operations.

There is a further requirement in the updating algorithm. After every
change of a
variable $\chi$ one has to recalculate the affected fields in
coordinate representation.
It is easy to get a similar estimate for the average number of
floating point operations
needed for such an updating. It just changes the coefficient of
$N_d$ from 8+4$d$ to 10+
4$d$.
  The ratio to the average number of floating point operations per site
   in a straightforward
  simulation in local simulation (coordinate representation) (6+2$d$)
  is then
  \begin{equation}
  \frac{t_w}{t_l}=\frac{10+4d}{6+2d}[n(2p-1)]^d[1+O(1/n)].
  \label{ratio}
  \end{equation}
  where $t_l$ and $t_w$ are the average times a sweep of the lattice
  takes in local
  and
  wavelet simulations, respectively.
  In particular, for Haar wavelets and two dimensions, $R_d\simeq
  (n^2/2)[1+O(1/n)]$.

  In fact, one can obtain a substantially smaller ratio as follows.
    Notice that the factor $(\log L)^d$ appears in (\ref{ratio})
  due to the computation of
  the coefficients of $\chi$ in action (\ref{action}). If one
  performs the
  Metropolis simulation in such a way that there are several
  hits at every
  wavelet $\chi$, then the same coefficients of (\ref{action})
  can be used and the extra calculation becomes
  not more costly then that of a hit in coordinate representation
  (local simulation).
   At the same time,
  having several hits at the same field component does not lead to
  any substantial
  savings in the local updating algorithm. Thus, in the limit of
  large number of hits
  the average sweeptimes for the two types of simulations converge.
  Of course,
  after a certain number of hits, the return in decreasing statistical
  errors
  diminishes. As usual, one has to find the optimal number of hits,
  so that the
  statistical errors were minimized using a fixed amount of
  computational time.
  Figure 1. shows the dependence of autocorrelation times on
  the number of hits.
  The autocorrelation times have been calculated using two
  different techniques.
  One, by direct measurement of the autocorrelation of the
  total magnetization,
  and the other by calculating the square of the relative
  error of the same
  physical quantity. They provide very similar information.
  When choosing the
  number of hits in the simulation, we had to keep in mind
  that the requirements
  on computing computing time slowly increase with the number
  of hits. The product
  of $\tau_w$ with the average time a sweep, $t_w$, should be
  optimized. We chose
  the optimum value of 5 in all of our simulations. For 5 hits
  we find the following
  values for $t_w/t_l$: 1.2, 1.3, 1.54, 1.82, on lattices of
  sizes $L=8,16,32,$
  and 64, respectively.

  The quality of an algorithm is determined by its improvement
  factor,
  which is defined as
  \begin{equation}
  I=\frac{t_w\tau_w}{t_l\tau_l},
  \label{improv}
  \end{equation}
  where $\tau_l$ and $\tau_w$ refer to real time autocorrelation
  lengths. Having
  determined (\ref{ratio}) in the rest
  of this paper we will concentrate on the determination of
  autocorrelation times
  $\tau_w$ and $\tau_l$.

  It is important to emphasize a substantial difference between
  a simulation using
  the standard, local (coordinate representation) algorithm and
  a simulation
  in wavelet representation. A hidden parameter of all simulations
  is the
  size of the window, $\Delta$, in which a new field value is
  picked during the
  Metropolis algorithm. In other words, if the value of the
  field component is
  $\phi$, then we choose a random new value for $\phi$, $\phi'$,
  in the interval $\phi-\Delta/2<\phi'
  <\phi+\Delta/2$. While in a local  simulation there is only one
  possible parameter
  $\Delta$, in the wavelet representation there is the possibility,
  and indeed the
  necessity, of choosing different values $\Delta_{ij}$,
  $i,j=1,...,n+1$,
  for different sizes of wavelets. The
  effect of this freedom of choice of the window sizes on
   autocorrelation times will be discussed later.

  The evaluation of the improvement factor $I$ of (\ref{improv})
  requires the determination
  of autocorrelation times.  We have performed extensive simulations
  on lattices of
  sizes $L=$8,16,32, and 64.
  In wavelet simulation, after a warmup period of $O(10^4)$ sweeps
  we read out
  the relevant physical variables in the next $O(10^5)$ sweeps.
  Using the local
  updating algorithm we had to have much longer runs, $O(10^6)$
  sweeps to be
  able to determine autocorrelation lengths. We compared the
  values of various
  physical quantities, such as $\langle\phi^2\rangle$,
  $\langle H\rangle$,
  $\langle |M|\rangle$, $\langle c_v\rangle$, and
  $\langle\chi\rangle$, where
  $M$ is the magnetization, $c_v$ is the specific heat, and
  $\chi$ is the
  magnetic susceptibility. All the physical quantities agreed within
  statistical errors in the two simulations. The simulations were
  performed at fixed $g=1$ and varying $m^2$ between the limits of
  -0.31 and -0.25, a range which includes the critical point.

  The purpose of the simulations was to calculate autocorrelation
  times, obtained
  from the $t$ dependence of autocorrelation functions
  \begin{equation}
  \Gamma(t)=\langle\ X(t')X(t'-t)\rangle_{t'},
  \label{gamma}
  \end{equation}
  where $X$ is a physical quantity, $t$ and $t'$ label real time during
  the simulation, measured in units of sweeps. The average of
  (\ref{gamma}) is
   over time $t'$. For moderate values of $t$, $\Gamma(t)$ is
   expected to
   have an exponential dependence on $t$,
   $\Gamma(t)\simeq \Gamma(0)\ e^{-t/\tau}$, where $\tau$ is
   the autocorrelation time.
   In fact, the integrated autocorrelation function,
   \begin{equation}
   \Gamma^I(t)=\frac{1}{\Gamma(0)}\sum_{t'=0}^t\Gamma(t')\simeq
    \frac{1}{1-e^{-1/\tau}}.
   \label{int}
   \end{equation}
   serves as a better measure of  the autocorrelation time
   ~\cite{sokal}~\cite{wolff}
   The integrated autocorrelation function goes into saturation
   at large $t$.
   The saturation value determines the autocorrelation time, $\tau$.

  We calculated autocorrelation
  times (measured in units of sweeps)
  for three physical quantities: Total magnetization, $M$,
  average value of $\phi(x)^2$, and the average value of the action.
    The dependence of autocorrelation times on the mass parameter of
  the Lagrangian, near the critical point is shown in Figure 2.
  Typical errors
  are about 5-10\%. The point of Fig. 2. represent data on
  lattice size $L=32$.

  It is important to determine the dependence of autocorrelation
  times on the
  size of the lattice.  Let us define an exponent $\zeta$ in a
  manner, somewhat
  analogous to (\ref{auto})
  \begin{equation}
  \tau=c L^\zeta .
  \label{auto2}
  \end{equation}
  At the critical point $\zeta$ coincides with $z$. Away from
  the critical point
  $\zeta$ is smaller then $z$ because $L>\xi$. The difference
  of $z$ and $\zeta$
  measures the finite size effect on the correlation length.
  The exponent $\zeta$,
  plotted in Fig. 3. in the critical region was calculated using
 all four lattice sizes we studied.
  Near the critical point, $m^2\simeq-0.285$, the exponent is
  almost 2. In fact,
  if the smallest lattice, $L=8$, is omitted from the fit the
  value exceeds 2. This
  shows that simulations in wavelet representation do not circumvent
  critical
  slowing down. In fact, a plot of exponent $\zeta$ in local simulations
  is almost identical to Fig. 3. In other words, the possible gain in
  autcorrelation times
  (\ref{auto2})
  in local and wavelet simulations. Indeed, the ratio of
  autocorrelation
  times in local and wavelet representation simulations has a
  large and fairly
  constant value for all lattice sizes considered. Fig. 4. shows
  these ratios in the
  critical region of $m^2$. A multiplier 5 corresponding to
  the number
of hits per wavelet coefficient
has been included in $\tau_w$.
 The ratio has only a moderate variation with lattice size
  or $m^2$. It is in the range of 10-15. Then the improvement
  factor, $I$,
   of (\ref{improv})
  at the critical point varies between 7 on the largest lattice to
  a value over 10 on the
  smaller lattices. There is a slight tendency to an increase of $I$
  when one moves
  away from critical point toward the disordered phase.

\section{Discussion}

The general expectation of substantially decreased autocorrelation times in
a simulation in wavelet representation was borne out by our simulations.
One can demonstrate quite dramatically the effectiveness of our method on
the free bosonic theory ($g=0$). Consider the autocorrelation of the total
magnetization. As we mentioned earlier, the magnetization is
proportional to one
of the wavelet expansion coefficients, the `non-wavelet' component. Let us
denote this component by $\chi$.
 Since the
kinetic term is  independent of $\chi$, $\chi$ completely decouples.
The $\chi$ dependence of the action
is
\begin{equation}
S=\frac{1}{2}m^2\chi^2.
\label{actionm}
\end{equation}
In other words, the action is a Gaussian function of $\chi$.
 Then the autocorrelation length for magnetization, can only depend
 on the dimensionless
 product $m\Delta$, where $\Delta$ is the range of change of $\chi$ in
 the simulation.
The dynamical critical exponent is 2. This can be seen by the
following heuristic
argument. Suppose $m\Delta<<1$. Then starting from an average value,
$\chi\sim 1/m$,
after $n$ sweeps the value of $\chi$ wanders away by an average value of
$\delta\chi\sim\sqrt{n}\Delta$ (random walk).
Complete decorrelation requires $\delta\chi\sim
\chi\sim1/m$. Then we obtain the following estimate for the
autocorrelation time:
\begin{equation}
\tau=n\simeq \frac{1}{(m\Delta)^2}=\frac{\lambda^2}{\Delta^2},
\label{auto4}
\end{equation}
where $\lambda$ is the spatial correlation length on the lattice.
(\ref{auto4})
shows not only that the critical exponent, $z=2$, but also
shows that the
numerical value of the autocorrelation length is inversely
proportional to
$\Delta^2$. In other words, if one chooses $\Delta$ for the
non-wavelet coefficient
too small one can get very long autocorrelation times. In a
wavelet simulation one
can control $\Delta$ independently for different types of
coefficients. Indeed,
in our simulation we have chosen the window $\Delta$ for
$\chi$ in such a  way
that the acceptance was approximately 0.5.

In a local, coordinate representation
simulation one has no control over the allowed range of change
of the total
magnetization. In fact, it is expected to change by a very
small amount in every
sweep. Since $\chi=\sum\phi/\sqrt{M}$, thus,
$\delta\chi=\sum\delta\phi/\sqrt{M}$,
roughly speaking $\delta\chi\sim\sqrt{M}\delta\phi/\sqrt{M}=\delta\phi$.
We performed some simulations at $g=0$ as well (free bosonic theory),
and in those simulations the optimal window $\Delta$ was independent of the
mass, $m$. Since the autocorrelation can only depend on the
product $m\Delta$, at small $m$ and constant $\Delta$ one can use
the asymptotic behavior
which should be the same as (\ref{auto4}). Indeed we obtained
the following
autocorrelation times for the total magnetization: 865 at
$m^2=.005$, 310 at
$m^2=.01$, 228 at $m^2=.015$, and 179 at $m^2=.02$. All of these
have about 10\%
error. These autocorrelation times give a good fit to (\ref{auto4}). In
other words, the effective value of $\Delta$ for the magnetization
was indeed  constant,
 as we assumed from the outset. Notice, however, what happens in a
simulation in wavelet representation. The window $\Delta$ is
adjusted for each type
of wavelet separately. In particular, one adjusts $\Delta$ for
the magnetization
as well. If one requires constant acceptance, then $\Delta m$ and so the
autocorrelation time is kept constant. The critical slowing down, at least for
free theories, is completely eliminated.

In the interacting case there is substantial mass renormalization.
Still, for moderate
values of the $m\Delta$ one
 expects a general dependence like (\ref{auto4}) on $\Delta$ where one
 should use the renormalized mass. Unfortunately, in interacting
 theories in $D=2$
 there is another dimensional quantity, $g$. The autocorrelation
 time also  depends
 on the dimensionless quantity $g/m^2$. The dependence on this
 quantity seems
 to dominate the behavior of exponent $\zeta$. The constant $c$ is,
 however, affected
 very strongly by the value of $m\Delta $, leading to a dramatic
 decrease of
 autocorrelation times in simulations in wavelet representation.

 To check the above ideas, we ran a series of simulations in
 wavelet representation
 on an $L=32$ lattice, at the same value of the physical parameters,
 set at the
 approximate location
 of the critical point, $g=1$ and $\mu^2=-.285$, but varying the
 window size for
 the wavelet component, that is proportional to the total
 magnetization. Fig. 5.
 shows
autocorrelation times for magnetization as a function of the
window size, $\Delta$,
for the appropriate wavelet component, $\chi=\sum\phi/\sqrt{M}$,
where $M$ is
the total number of lattice sites. The curve has, for these values of
$m^2$ and $g$,
 a broad minimum around an optimal
window size.

The  optimal value of window size for the
change of $\phi$ in local simulations is about $\Delta=3$. Due to
the orthogonality
of the transformation from local variables, $\phi$, to wavelet
components, $\chi$,
a local simulation is in some sense equivalent to a wavelet
simulation in which
every window $\Delta$ is set equal to the value of $\Delta$ in
the local simulation.
For this reason we ran a simulation on a 32$\times$32 lattice
at $m^2=-.285$,
setting all windows equal to $\Delta=3.0$. We obtained an
autocorrelation time of
$\tau_w=226$, comparable to $\tau_l\simeq436$. At the same point,
if we optimize all the windows, such that all acceptances are
between 0.4 and 0.5,
then $\tau_w=39$. This result supports the idea that a
major part of the decrease in autocorrelation time
is not the result of
relaxing the system at different scales simultaneously. It is rather
the consequence of the extra freedom of setting the windows for every
scale independently.

In the present paper we have used Haar wavelets. It would be of
some iterest to
investigate simulations with other examples of the Daubechies
wavelet series.
Because the wavelet expansion coefficients would be all
different, the calculation
of the action would take a longer time. It is still
possible that the improved
convergence properties of the wavelet expansion would
make the comparision with
the local simulation even more favorable.

Finally, we would like to point out that wavelet representations
can be used for
theories with more complicated order parameters as well. In such
simulations
the ratio of sweeptimes should be roughly the same as has been
obtained in the
current paper. As an example, for compact  Abelian gauge theories
wavelets should
be formed from the components of the vector potential.
The calculation of the
change of the action requires the calculation of trigonometric function in
both representations. The estimate we have given for computational
requirements in Sec. 2. would still stand. Unfortunately, since the action
is nonpolynomial, the ratio of simulation times may not be
substantially improved by repeated
hits at individual wavelet coefficients.

\section*{ACKNOWLEDGEMENTS}

The work of P. S. is supported by funds from Research Committee of the
Higher Education Funding Council for Wales (HEFCWR).
  He is also indebted for the partial support of
the United States Department of Energy under
grant no.  DE-FG02-84ER40153.

\newpage
\begin{center}
{\Large FIGURE CAPTIONS}
\end{center}
\begin{flushleft}
Fig. 1.
\end{flushleft}
\begin{quote}
$\tau_w$ for magnetization as a function of the number of hits per
wavelet coefficient on a 32$\times32$ lattice. The diamonds
represent a direct
measurement, while the triangles represent $\tau_{\rm eff}$
calculated from
relative errors.
\end{quote}
\begin{flushleft}
Fig. 2.
\end{flushleft}
\begin{quote}
$\tau_w$ as a function of $m^2$ for the total magnetization (diamonds),
for $\langle\phi^2\rangle$ (triangles), and for the total energy (stars).
\end{quote}
\begin{flushleft}
Fig. 3.
\end{flushleft}
\begin{quote}
The exponent $\zeta$ as a function of $m^2$.
\end{quote}
\begin{flushleft}
Fig. 4.
\end{flushleft}
\begin{quote}
The ratio of autocorrelation times $\tau_l/\tau_w$ as a function of $m^2$.
Figures 4a, 4b, 4c, and 4d correspond to lattices of size
$L=$8,16,32, qnd 64,
respectively.
\end{quote}
\begin{flushleft}
Fig. 5.
\end{flushleft}
\begin{quote}
Autocorrelation time $\tau_w$ as a function of $\Delta$ at $m^2=-.285$ on
a lattice of size $L=32$.
\end{quote}

 \end{document}